\documentclass[a4paper,12pt]{article}
\begin{document}

\title{NPT Bound Entanglement- The Problem Revisited}

\author{Indrani Chattopadhyay\thanks{ichattopadhyay@yahoo.co.in}\ \ and Debasis Sarkar
\thanks{dsappmath@caluniv.ac.in, debasis1x@yahoo.co.in}}
\date{}
\maketitle

\begin{center}
Department of Applied Mathematics, University of Calcutta,\\ 92,
A.P.C. Road, Kolkata- 700009, India.
\end{center}

\begin{abstract}
Recently [quant-ph/0608250] again created a lot of interest to prove
the existence of bound entangled states with negative partial
transpose (NPT) in any $d \times d (d \geq 3)$ Hilbert space.
However the proof in quant-ph/0608250 is not complete but it shows
some interesting properties of the Schmidt rank two states. In this
work we are trying to probe the problem in a different angle
considering the work by D\"{u}r et.al [Phys. Rev. A, 61,
062313(2000)]. We have assumed that the Schmidt rank two states
should satisfy some bounds. Under some assumptions with these bounds
one could prove the existence of NPT bound entangled states. We
particularly discuss the case of two copy undistillability of the
conjectured family of NPT states. Obviously the class of NPT bound
entangled states belong to the class of conjectured to be bound
entangled states by Divincenzo et.al [Phys. Rev. A, 61,
062312(2000)] and by D\"{u}r et.al [Phys. Rev. A, 61, 062313(2000)].
However the problem of existence of NPT bound entangled states still
remain open.

PACS number(s): 03.67.Hk, 03.65.Ud.

\end{abstract}
\maketitle \vspace{.5cm}

\section{Introduction}

The basic issue on the classification of mixed state entanglement at
least on the level of bipartite systems solely depends upon whether
there exist bound entangled states or not. The existence of
PPT-bound (PPT means positive partial transpose) entangled
states~\cite{horodecki98} and also the existence of NPT  $N-$copy
undistillable states~\cite{div,dur} for every positive integer $N$
naturally indicates there may exist NPT-bound entangled states.
Recent work ~\cite {simon} also indicates positively the existence
of NPT- bound entangled states. In this work we consider the problem
of the existence of NPT-bound entangled states with some assumptions
on Schmidt rank two states. We first briefly describe the issue and
the importance of the problem.

In recent years it is found that quantum entanglement is an useful
resource in performing several tasks in quantum information theory
and quantum communication~\cite{entanglement}. Maximally entangled
states shared between two parties are essential ingredients in this
respect~\cite{mes}. Now due to the interaction with environment,
states are in practice found to be mixed. However there is a process
called distillation by which we can distill sometimes maximally
entangled states out of certain pair of mixed entangled states using
only local operation and classical communications
(LOCC)~\cite{locc}. But entanglement of a state is not always
sufficient for distillability. The Peres-Horodecki
criterion~\cite{peres-hor} namely the partial transpose
corresponding to any bipartite system gave us a necessary condition
for distillability of any entangled state. If a bipartite density
operator have positive partial transpose then it is not distillable.
Further any PPT-state may be classified into two classes, separable
and PPT-bound entangled states (bound entangled states means no
entanglement can be extracted from them by LOCC, i.e., not
distillable). There exist PPT-bound entangled
states~\cite{horodecki98}. But are all NPT- states which are
necessarily entangled, distillable~\cite{lew}? Until this time there
is no answer. Independently, Divincenzo et.al~\cite{div} and D\"{u}r
et.al~\cite{dur} and also Somshubhro et.al~\cite{som} gave some
evidence for $N-$copy undistillable states. Watrous \cite{john}
further investigated the problem of distillability with large number
of copies of some entangled states. Recently the work by
Simon~\cite{simon} indicates there may exist NPT bound entangled
states with a large class of state that includes the conjectured
family of NPT bound entangled states. Here we show that NPT-bound
entangled states exist for any bipartite $d \times d (d \geq 3)$
system if we consider some simple assumptions on Schmidt rank two
states. Our approach is based on some bounds of rank two states that
are not closed in \cite{dur}. Obviously they belong to the classes
as suggested earlier. Lastly we show a simple property that
satisfied by a class of conjectured bound entangled states.

With the existence of NPT-bound entangled state it is also proved
that the distillable entanglement is nonadditive and not
convex~\cite{shor}. It should be noted that by distillable
entanglement~\cite{locc,dist} of a bipartite state we mean how much
pure maximally entangled states we can extract asymptotically by
means of LOCC from several copies of that state. Now, by definition
of bound entanglement, every bound entangled state, whether NPT or
PPT, has zero distillable entanglement. In~\cite{shor}, Shor et. al
showed that distillable entanglement of tensor product of two
states, one PPT-bound entangled state (formed by pyramid UPB) and
another conjectured to be NPT-bound entangled state(which we shall
prove really NPT-bound entangled) is non-zero. Which proves the
nonadditivity and non-convexity of distillable entanglement. Also,
it constitutes another example that PPT-bound entangled states can
be used in the activation process~\cite{shor,activ,activ1}.

\section{The conjectured class of NPT bound entangled states}

Before going to discuss our result, we first mention the notion of
distillable states on any bipartite system described by the joint
Hilbert space $H_A \otimes H_B$.

\noindent {\em Definition}~\cite{div,dur,shor}.-- A density matrix
$\rho$ is distillable if and only if there exists a positive integer
$n$ such that

\begin{equation}
\langle \psi | (\rho^{\otimes n})^{T_A} |\psi \rangle < 0
\end{equation}
for any Schmidt rank two state $|\psi \rangle \in (H_A \otimes
H_B)^{\otimes n}$, where $T_A$ represents partial transpose with
respect to the system $A$.

Now we consider the key state in $d\times d (d\geq 3)$ Werner class
that are one copy undistillable ~\cite{div,dur,shor} and conjectured
to be bound entangled.

\noindent {\em Theorem/Conjecture}.-- The state $\rho (\lambda )$ in
$d\times d, \ (d\geq 3)$ of Werner class represented by
\begin{equation}
\rho (\lambda ) = \frac{1}{d(d+ \lambda (d-1))}[I + \lambda
\sum_{i,j, i<j} P(|ij \rangle - |ji \rangle )],
\end{equation}
where $\frac{1}{d-1} < \lambda \leq 1$ is NPT- bound entangled,
where, $\{|0 \rangle, |1 \rangle, |2 \rangle \cdots \}$ is an
orthonormal basis on the Hilbert space $H_A (H_B)$.

For $\lambda =\frac{1}{d-1}$ the state is separable.

The state is one copy undistillable~\cite{div,dur,shor}. One has to
prove, it is $n$-copy undistillable for any $n$.

\subsection{Two copy undistillability under some assumptions}

To explain the possibility of two copy undistillability of the
conjectured class, we consider first the partial transpose of the
given state $\rho(\lambda )$. The partial transpose of the state
with respect to the system $A$ is,

$$\rho(\lambda )^{T_A} =
\frac{1}{d(d+ \lambda (d-1))}[(1+ \lambda)I - \lambda P(\sum_{i} |ii
\rangle )].$$ In the sequel we write $P^+ = P(\sum_{i} |ii \rangle
).$ Before mentioning our basic assumptions which seems to be
correct for any Schmidt rank two states, we first note an
observation found for a class of Schmidt rank two states.

 \noindent {\em An Observation}.-- Since any Schmidt rank two state
$|\psi \rangle$ in $(H_A \otimes H_B)^{\otimes 2}$ has expectation
value less than two with the operator $I_{AB} \otimes P^+_{AB} $,
therefore the rank two states of the form $ |\chi_A \phi_B \rangle
\otimes |\psi^{\prime} \rangle_{AB}$, where
$|\psi^{\prime}\rangle_{AB}$ is any Schmidt rank two state in $(H_A
\otimes H_B)$, has expectation value less than three with the
operator $I_{AB} \otimes P^+_{AB} + P^+_{AB} \otimes I_{AB}$.

(in the text we have used multiple copies of operators of system
$A,B$ with the same suffix)

With this simple observation on Schmidt rank two states, if someone
ask what will be the case for any general Schmidt rank two states in
$(H_A \otimes H_B)^{\otimes 2}$? We mention this as our first
assumption.

\noindent {\em Assumption 1}.-- For any Schmidt rank two states
$|\psi \rangle$ in $(H_A \otimes H_B)^{\otimes 2}$,
\begin{equation}
 \langle \psi
|(kI_{AB} - P^+_{AB})\otimes P^+_{AB} + P^+_{AB} \otimes (kI_{AB} -
P^+_{AB})|\psi \rangle \leq \max \{2k,~ 3k-4 \},
\end{equation}
where $k>2$.

 Clearly, with this bound it is now easy to check that
for any Schmidt rank two state $|\psi \rangle$ in $(H_A \otimes
H_B)^{\otimes 2}$,
\begin{equation}
\langle \psi | (\frac{k}{2}I_{AB} - P^+_{AB})^{\otimes 2} |\psi
\rangle \geq \frac{k^2}{4} - \frac{\max \{2k,~ 3k-4 \}}{2}\geq 0,
\verb"if" ~~ k \geq 4
\end{equation}

Now, putting $k= \frac{2(1+ \lambda )}{\lambda}$, we have,
$$\langle \psi | (\rho(\lambda)^{\otimes
2})^{T_A} |\psi \rangle \geq 0, ~~\verb"for" ~~\lambda \leq 1$$
 i.e., $\rho(\lambda)$ is two copy undistillable.

\subsection{n-copy undistillability under some assumptions}
Next we consider some assumptions in $(H_A \otimes H_B)^{\otimes
n}$, for any $n \geq 2$. We have assumed a sequence of bounds for
any Schmidt rank two state $|\psi \rangle$ in $(H_A \otimes
H_B)^{\otimes n}$ where $n \geq 2$. These bounds are not present in
\cite{dur}, however from numerical evidences and also analytically
for large classes of Schmidt rank two states we found these are
true. Until now we find no exceptional cases that violets these
bounds.

$$(i).~~\langle \psi |(I_{AB})^{\otimes (n-1)} \otimes
P^+_{AB} + \cdots + P^+_{AB} \otimes (I_{AB})^{\otimes (n-1)}|\psi
\rangle \leq 2(C^n_1 -C^{n-1}_1)+ C^{n-1}_1 = n+1, $$

$$(ii).~~\langle \psi |(I_{AB})^{\otimes (n-2)} \otimes
(P^+_{AB})^{\otimes 2} + \cdots + P^+_{AB} \otimes (I_{AB})^{\otimes
(n-2)} \otimes P^+_{AB} + \cdots  $$

$$+ (P^+_{AB})^{\otimes 2}
\otimes (I_{AB})^{\otimes (n-2)}|\psi \rangle \leq 2(C^n_2
-C^{n-1}_2)+ C^{n-1}_2 = 2(C^{n-1}_1)+ C^{n-1}_2,$$

$$(iii).~~\langle \psi |(I_{AB})^{\otimes (n-3)} \otimes
(P^+_{AB})^{\otimes 3} + \cdots + P^+_{AB} \otimes (I_{AB})^{\otimes
(n-3)} \otimes (P^+_{AB})^{\otimes 2} + \cdots $$

$$ + (P^+_{AB})^{\otimes 3} \otimes (I_{AB})^{\otimes (n-3)}|\psi
\rangle \leq 2(C^n_3 -C^{n-1}_3)+ C^{n-1}_3 = 2(C^{n-1}_2)+
C^{n-1}_3,$$

Proceeding in this way we have for any $m<n+1$,
$$(m).~~\langle \psi |(I_{AB})^{\otimes n-m} \otimes
(P^+_{AB})^{\otimes m} + \cdots + P^+_{AB} \otimes (I_{AB})^{\otimes
n-m} \otimes (P^+_{AB})^{\otimes m-1} + \cdots$$

$$ + (P^+_{AB})^{\otimes m} \otimes (I_{AB})^{\otimes n-m}|\psi
\rangle \leq 2(C^n_m -C^{n-1}_m)+ C^{n-1}_m = 2(C^{n-1}_{m-1})+
C^{n-1}_m,$$ where $C^n_r =
\frac{n.(n-1)...(n-r+1)}{r.(r-1)....2.1}$, for any $r \leq n$.

Now consider the set of all Schmidt rank two states $|\psi \rangle$
in $(H_A \otimes H_B)^{\otimes n}$ that attains the optimal values
for all the above bounds. The set is not empty. For such a rank two
state $|\psi \rangle$ we have,
\begin{equation}
\langle \psi | (\rho(\lambda)^{\otimes n})^{T_A} |\psi \rangle \geq
0.
\end{equation}

Now we conjecture that the above result satisfied by any rank two
state from the optimal set is also satisfied by any rank two state.
In other words we want to say that to prove  $\rho(\lambda)$ is
$n$-copy undistillable for any $n$, the proof for optimal class of
states is sufficient. However one may construct some composite
bounds like equation (3) that would directly prove the $n$-copy
undistillability of the conjectured class of Werner states
represented by $\rho(\lambda)$. We have tested for a large class of
rank two states that satisfied by equation(5). Also it is
interesting to note that if the conjectured class of states are NPT
bound entangled states, then the bounds we have assumed must be
satisfied for Schmidt rank two states. Next we consider a simple
property satisfied by the conjectured class of bound entangled
states.

\section{A simple property}
For simplicity we consider the state $\rho(\lambda )$ for $\lambda
=1$. We denote it by $\rho$. First, we consider $\rho^{\otimes 2}$.
After rewriting it with first two basis elements for system $A$ then
for system $B$, and omitting normalization factor (as it will not
alter the trace condition) it looks as follows:

$$\rho^{\otimes 2} = I\otimes I + \sum_{i,j,m,n, i<j}[ P(|minj \rangle - |mjni
\rangle) + P(|imjn \rangle - |jmin \rangle)]$$
\begin{equation}
+\sum_{i,j,k,l, i<j, k<l}P(|ikjl \rangle - |iljk \rangle - |jkil
\rangle + |jlik \rangle)
\end{equation}
$$ = I\otimes I + \sum_{i,j,m, i<j}[ P(|mimj \rangle - |mjmi
\rangle) + P(|imjm \rangle - |jmim \rangle)] $$

$$ + \sum_{i,j,k,l, i<j, k<l}[ P(|ikjl \rangle - |iljk \rangle) +
P(|ikjl \rangle - |jkil \rangle)$$ \hspace{2cm}$+ P(|ikjl \rangle -
|iljk \rangle - |jkil \rangle + |jlik \rangle)],$

where $i,j,k,l,m,n = 0,1,2,....$.

\vspace{.5cm} Now it is easy to check that any off-diagonal operator
of the form,\\
$ |ijkl \rangle \langle mnpq|, ~~ i,j,k,l,m,n,p,q = 0,1,2,...$,
occurs maximum once or twice on the above expression. After taking
partial transpose say with respect to system $A$ it takes the form $
|mnkl \rangle \langle ijpq|,$ $i,j,k,l,m,n,p,q = 0,1,2,...$. Also if
we take the partial transpose of $\rho^{\otimes 2}$ with respect to
system $A$, then we find in $(\rho^{\otimes 2})^{T_A}$, $P(|ijpq
\rangle), P(|mnkl \rangle), i,j,k,l,m,n,p,q =0,1,2,...,$ occur same
times as $|mnkl \rangle \langle ijpq|$, $i,j,k,l,m,n,p,q =
0,1,2,...$. Therefore, trace with any Schmidt-rank two state $|\psi
\rangle$ in the basis we have represented $(\rho^{\otimes
2})^{T_A}$, would be always non-negative.

Next consider $\rho^{\otimes 3}$. It will take the form, if we write
first three basis elements for system $A$ and then for system $B$,
as follows:

$$\rho^{\otimes 3} = I\otimes I\otimes I + \sum_{i,j,m,n,p,q, i<j}[ P(|mpinqj \rangle
- |mpjnqi \rangle)$$ $ +  P(|mipnjq \rangle - |mjpniq \rangle) +
P(|impjnq \rangle - |jmpinq \rangle)]$
$$+\sum_{i,j,k,l,m,n, i<j, k<l} [P(|miknjl \rangle - |milnjk \rangle
- |mjknil \rangle $$
$$+ |mjlnik \rangle) + P(|imkjnl \rangle -
|imljnk \rangle - |jmkinl \rangle + |jmlink \rangle)$$
$$ +
P(|ikmjln \rangle - |ilmjkn \rangle - |jkmiln \rangle + |jlmikn
\rangle)]$$
$$+\sum_{i,j,k,l,m,n, i<j, k<l, m<n}P(|ikmjln \rangle - |iknjlm
\rangle - |ilmjkn \rangle$$
\begin{equation}
 + |ilnjkm \rangle - |jkmiln \rangle +
|jknilm \rangle + |jlmikn \rangle - |jlnikm \rangle)
\end{equation}

Here again if we consider any off-diagonal operator of the form,\\
$|ijklmn \rangle \langle pqrstu|,~~ i,j,k,l,m,n,p,q,r,s,t,u =
0,1,2,...$, occur maximum one or $2^1$ or $2^2$ times on the above
expression. Therefore, with the similar argument above, we find
trace with any Schmidt-rank two state $|\psi \rangle$ in the basis
we have represented $(\rho^{\otimes 3})^{T_A}$, would be always
non-negative.

For $\rho^{\otimes N}$, if we write it similarly as above, we find
off-diagonal operator of the form $$ |ijk \cdots N \ {\rm times} \
lmn \cdots N \ {\rm times} \ \rangle \langle pqr \cdots N \ {\rm
times}~~ stu \cdots N \ {\rm times}\ |,$$ $i,j,k,l,m,n,p,q,r,s,t,u =
0,1,2,...$, will occur maximum one or $2^1$ or $2^2$ $\cdots$ or
$2^{N-1}$ times and in the partial transpose $(\rho^{\otimes
N})^{T_A}$, \\
$P(|pqr \cdots N \ {\rm times} \ lmn \cdots N \ {\rm times} \
\rangle)$, $P(|ijk \cdots N \ {\rm times} \ stu \cdots N \ {\rm
times}\rangle)$, occur same number of times as the off-diagonal
operator, $$ |pqr \cdots N \ {\rm times} \ lmn \cdots N \ {\rm
times} \ \rangle \langle ijk \cdots N \ {\rm times}~~ stu \cdots N \
{\rm times}\ |.$$ So the trace with any Schmidt rank two state
$|\psi \rangle$ in the basis we have considered $(\rho^{\otimes
N})^{T_A}$, would be always non-negative. However for any Schmidt
rank two state (i.e., in any other basis), we are unable to
calculate the trace with $(\rho^{\otimes N})^{T_A}$ for any $N$,
using the property we have found above. The property we have
discussed above for $\rho$, could be easily extended to any $\rho
(\lambda )$.

\section{Conclusion}

To summarize our results, we have revisited the problem of existence
of NPT-bound entangled states of any bipartite systems $d\times d, d
\geq 3$ with some assumptions made on Schmidt rank two states. The
key role plays here the bounds that we have assumed for any rank two
states. Their proof would readily solve the problem of
classification of states at least at bipartite level, i.e., whether
a state is either separable or bound entangled (PPT or NPT) or
distillable. There are always some confusion regarding
distillability when a bipartite state is NPT. However the problem of
existence of NPT bound entangled states remains still open.

 \vspace{1cm}
{\bf Comment.} We have started the problem since the year, late
2003. Sometimes we felt, we have solved the problem, after that we
found the proof is not complete yet. Recently (last 8-10 months) we
found some bounds by which the problem can be solved. We found them
through some requirements of some operators to maintain positivity.
However that proof is not most general one. Then we look upon the
problem in reverse order using those bounds and found some
assumptions that we have mentioned in our first version. After
observing the paper quant-ph/0608250, we put our work into the net
also. Since the assumption made in our first version is strong
enough and need not to be satisfied for all rank two states,
therefore we dropped this assumption in second version, which is not
actually needed for our proof. In this version we have tried to be
more explanatory and also discusses the problem in different angles.

 {\bf Acknowledgement.} The authors thank K.G. Vollbrecht for his valuable
comments and suggestions. D.S. thanks G. Kar, R. Simon, S.Ghosh
regarding the issue of this problem. I.C. acknowledges CSIR, India
for providing fellowship during this work.

\end{document}